\documentclass[conference]{IEEEtran}
\IEEEoverridecommandlockouts
\usepackage{cite}
\usepackage{amsmath,amssymb,amsfonts}
\usepackage{algorithmic}
\usepackage{graphicx}
\usepackage{textcomp}
\usepackage{xcolor}

\usepackage{booktabs} 

\usepackage[english]{babel}
\usepackage{moresize}
\usepackage{amsmath}

\usepackage{algorithmic} 
\usepackage{balance}
\usepackage{comment}
\usepackage{paralist}
\usepackage{multirow}

\usepackage{filecontents}

\usepackage[english]{babel}
\usepackage[latin1]{inputenc}
\usepackage{mathrsfs}
\usepackage{graphicx}
\usepackage{amssymb}
\usepackage{url}
\usepackage{subfigure}
\usepackage{amsmath}
\usepackage{enumitem}
\usepackage[linesnumbered,algoruled,boxed,lined]{algorithm2e}
\usepackage{adjustbox}
\usepackage{amssymb}
\usepackage{times}
\usepackage{hyperref}

\usepackage{pgfplots}
\usetikzlibrary{pgfplots.dateplot}

\usepackage{filecontents}
\definecolor{tblue}{RGB}{31,119,180}
\definecolor{torange}{RGB}{255,127,14}
\definecolor{tgreen}{RGB}{44,160,44}
\definecolor{tred}{RGB}{214,39,40}
\definecolor{tpurple}{RGB}{148,103,189}

\usepackage{filecontents}

\newcommand{\hide}[1]{} 

\newcommand{\ie}{\textit{i}.\textit{e}.}
\newcommand{\eg}{\textit{e}.\textit{g}.}

\def\BibTeX{{\rm B\kern-.05em{\sc i\kern-.025em b}\kern-.08em
    T\kern-.1667em\lower.7ex\hbox{E}\kern-.125emX}}
    
\begin{document}


\title{Multi-Behavior Enhanced Recommendation with Cross-Interaction Collaborative Relation Modeling

\thanks{\textbf{*Corresponding author: Yong Xu.}}}

\def\model{GNMR}
\def\full{}



\author{\IEEEauthorblockN{Lianghao Xia$^1$, Chao Huang$^2$, Yong Xu$^{1*}$, Peng Dai$^2$, Mengyin Lu$^2$, Liefeng Bo$^2$}
\IEEEauthorblockA{South China University of Technology$^1$, JD Finance America Corporation$^2$\\
cslianghao.xia@mail.scut.edu.cn, yxu@scut.edu.cn, chaohuang75@gmail.com, \{peng.dai,mengyin.lu,liefeng.bo\}@jd.com
}}


\maketitle

\begin{abstract}

Many previous studies aim to augment collaborative filtering with deep neural network techniques, so as to achieve better recommendation performance. However, most existing deep learning-based recommender systems are designed for modeling singular type of user-item interaction behavior, which can hardly distill the heterogeneous relations between user and item. In practical recommendation scenarios, there exist multi-typed user behaviors, such as browse and purchase. Due to the overlook of user's multi-behavioral patterns over different items, existing recommendation methods are insufficient to capture heterogeneous collaborative signals from user multi-behavior data. Inspired by the strength of graph neural networks for structured data modeling, this work proposes a \underline{\textbf{G}}raph \underline{\textbf{N}}eural \underline{\textbf{M}}ulti-Behavior Enhanced \underline{\textbf{R}}ecommendation (\model) framework which explicitly models the dependencies between different types of user-item interactions under a graph-based message passing architecture. \model\ devises a relation aggregation network to model interaction heterogeneity, and recursively performs embedding propagation between neighboring nodes over the user-item interaction graph. Experiments on real-world recommendation datasets show that our \model\ consistently outperforms state-of-the-art methods. The source code is available at https://github.com/akaxlh/GNMR.\\
\end{abstract}


\section{Introduction}
\label{sec:intro}

Recommender systems become the essential part of online platforms, to alleviate the information overload problem and make recommendation for users~\cite{shi2018heterogeneous,huang2019online}. The key objective of recommendation frameworks is to accurately capture user's preference over different items based on their observed interactions~\cite{wang2019neural,xu2020social}. As effective feature learning paradigms, deep learning has attracted a lot of attention in recommendation, which results in various neural network-based methods being proposed for user-item interaction modeling~\cite{wang2020disentangled,du2018collaborative,zheng2016neural,he2017neural}. These methods transform users and items to vectorized representations based on different neural network structures. For example, autoencoder has been used for latent representation projection in collaborative filtering~\cite{wu2016collaborative}. To endow the collaborative filtering architecture with the capability of non-linear feature interaction modeling, NCF~\cite{he2017neural} integrates the matrix factorization and multi-layer perceptron network.



However, these recommendation solutions mostly focus on singular type of user-item interaction behavior. In real-world online applications, users' behaviors are multi-typed in nature, which involves heterogeneous relations (\eg, browse, rating, purchase) between user and item~\cite{xia2020multiplex,guo2019buying}. Each type of user behavior may exhibit different semantic information for characterizing diversified interactive patterns between user and item. Hence, the current encoding functions of user-item interactions are insufficient to comprehensively learn the complex user's preference. While having realized the importance of leveraging different types of user behaviors, encoding multi-typed behavioral patterns present unique challenges which cannot be easily handled by recommendation methods designed for single type of user-item interactions. In particular, it is non-trivial to effectively capture implicit relationships among various types of user behaviors. Such different types of interaction behaviors may be correlated in complex ways to provide complementary information for learning user interests. Additionally, although there exist some recent developed multi-behavior user modeling techniques for recommendation~\cite{xia2020multiplex,gao2019neural}, they fail to capture the high-order collaborative effects with the awareness of different user-item relations. Taking the inspiration from the effectiveness by employing graph neural networks in recommendation~\cite{wu2019reviews,wang2019neural}, it is beneficial to consider high-order relations between user-item interaction into the embedding space. \\\vspace{-0.1in}

\noindent \textbf{Contribution}. In this work, we propose a \underline{\textbf{G}}raph \underline{\textbf{N}}eural \underline{\textbf{M}}ulti-Behavior Enhanced \underline{\textbf{R}}ecommendation framework (short for \model), to capture users' preference on different items via a multi-behavior modeling architecture. Specifically, the designed graph neural multi-behavior learning framework explores high-order user-item interaction subgraph, characterizing complex relations between different types of user behaviors in an automatic manner. In our graph neural network, we design a relation dependency encoder to capture the implicit dependencies among different types of user behaviors under a message passing architecture. With the aim of modeling the graph-structured interactions between users and items, our developed \model\ performs the embedding propagation over the multi-behavior interaction graph in a recursive way, with the injection of type-specific behavior relationships. We evaluate our framework on real-world datasets from MovieLens, Yelp and Taobao. Evaluation results show the effectiveness of our \model\ model compared with state-of-the-art baselines.

The main contributions of this work are summarized as:

\begin{itemize}[leftmargin=*]
\item This work focuses on capturing behavior type-aware collaborative signals with the awareness of high-order relations over user-item interaction graph in the embedding paradigm for recommendation.

\item We propose a new graph neural network framework \model\ for multi-behavior enhanced recommendation, with the exploration of dependencies between different types of behaviors under a message passing architecture. \model\ performs the embedding propagation between users and items on their graph-structured connections, and aggregates the underlying cross-type behavior relations with a high-order scenario.

\item Experimental results on three real-world datasets from online platforms show the effectiveness of our \model\ model compared with state-of-the-art baselines.
\end{itemize}

\section{Preliminaries}
\label{sec:model}

In a typical recommender system, there exist two types of entities, \ie, set of users $U$ ($u_i\in U$) and items $V$ ($v_j\in V$), where $|U|=I$ (indexed by $i$) and $|V|=J$ (indexed by $j$). In our multi-behavior enhanced recommendation scenario, users could interact with items under multiple types of interactions (\eg, different ratings, browse, purchase). We define a three-dimensional tensor $\textbf{X} \in \mathbb{R}^{I\times J\times K}$ to represent the multi-typed user-item interactions, where $K$ denotes the number of behavior types (indexed by $k$). In tensor $\textbf{X}$, each element $x_{i,j}^k=1$ if there exist interactions between user $u_i$ and item $v_j$ given the $k$-th behavior type and $x_{i,j}^k=0$ otherwise. In this work, we aim to improve the recommendation performance on the target behavior of users, by exploring the influences of other types of behaviors. We formally define the problem as:\\\vspace{-0.12in}

\noindent \textbf{Problem Statement}. Given the multi-behavior interaction $\textbf{X} \in \mathbb{R}^{I\times J\times K}$ under $K$ behavior types, the goal is to recommended a ranked list of items in terms of probabilities that user $u_i$ will interact with them under the target behavior type.\\\vspace{-0.12in}

\noindent \textbf{Connections to Existing Work}. Recent years have witnessed the promising results of graph neural networks in learning dependence from graph-structured data~\cite{wu2020comprehensive}. In general, the core of graph neural networks is to aggregate feature information from adjacent vertices over a graph under a message passing architecture~\cite{ying2018graph}. Inspired by the effectiveness of graph convolutional network, recent studies, such as NGCF~\cite{wang2019neural} and GraphSage~\cite{hamilton2017inductive}, explore user-item interaction graph to aggregate the embeddings from neighboring nodes by employing graph convolutional network in collaborative filtering. These work performs information propagation between vertices to exploit relations among users and items. This work extends the exploration of designing graph neural network models for recommendation by tackling the challenges in dealing with multi-typed user-item interactions for recommendation.
\section{Methodology}
\label{sec:solution}

\begin{figure}
    \centering
    \includegraphics[width=1.0\columnwidth]{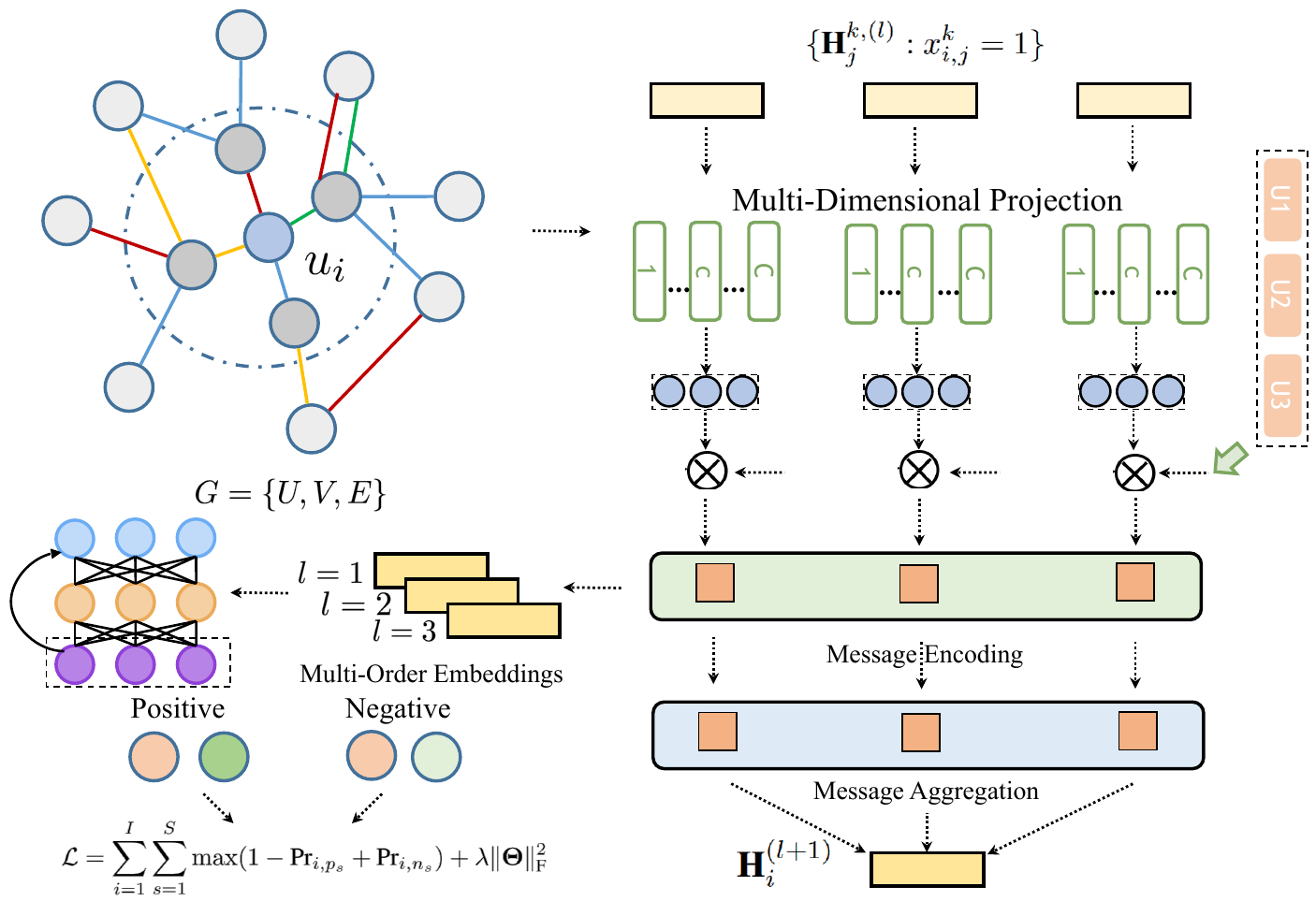}
    \vspace{-0.2in}
    \caption{Model Architecture of \model.}
    \label{fig:framework1}
    \vspace{-0.2in}
\end{figure}

This section presents the details of our proposed neural framework \model\ for multi-behavior enhanced recommendation. Based on the preliminaries in Section~\ref{sec:model}, we generate the user-item interaction graph $G=\{U,V,E\}$, where nodes are constructed by user set $U$ and item set $V$. $E$ represents the edges connecting users and items based on their multi-typed interactions. Specifically, in graph $G$, there exists an edge $e_{i,j}^k$ between user $u_i \in U$ and item $v_j \in V$ under the $k$-th behavior type if $x_{i,j}^k=1$. The general graph neural network utilizes $G$ as the computation graph for information diffusion, during which the neighborhood messages are aggregated to obtain contextual representations~\cite{hu2020heterogeneous,zhang2019heterogeneous}. In our \model\ framework, we update user and item latent embeddings by propagating them on the user-item multi-behavior interaction graph $G$, to capture the type-aware behavior collaborative signals for recommendation. We show the model architecture of \model\ in Figure~\ref{fig:framework1}.
\vspace{-0.05in}

\subsection{Type-specific Behavior Embedding Layer}
We suppose the output embedding of the $l$-th \model\ graph layer as $\textbf{H}^l$ which is then fed into the $(l+1)$-th layer as input representation. Given edges between users and items under the behavior type of $k$, we construct the passed message as below:
\begin{align}
\textbf{H}_{i\leftarrow}^{k,(l)} =  \eta(\{\textbf{H}_{j}^{k,(l)}:x_{i,j}^k=1\})\nonumber\\
\textbf{H}_{j\leftarrow}^{k,(l)} =  \eta(\{\textbf{H}_{i}^{k,(l)}:x_{i,j}^k=1\})
\end{align}
\noindent where $\textbf{H}_{i\leftarrow}^{k,(l+1)} \in \mathbb{R}^d$ and $\textbf{H}_{j\leftarrow}^{k,(l+1)} \in \mathbb{R}^d$ denotes the embeddings passed to $u_i$ and $v_j$, respectively. $\eta(\cdot)$ represents the embedding layer which preserves the unique characteristics of each type (\ie, $k$) of user behaviors. During the message propagation process, we initialize the embeddings $\textbf{H}_{i}^{0}$ and $\textbf{H}_{j}^{0}$ for user $u_i$ and item $v_j$ by leveraging Autoencoder-based pre-training scheme~\cite{sedhain2015autorec} for generating low-dimensional representations based on multi-behavior interaction tensor $\textbf{X}$.

In our message passing architecture, $\eta(\cdot)$ is designed to obtain representations for each behavior type of $k$, by considering type-specific behavior contextual signals (\eg, behavior frequency). We formally represent $\eta(\cdot)$ as follows:
\begin{align}
\alpha_{c,k} = \delta(\sum_{j\in N(i,k)} \textbf{W}_1 \cdot \textbf{H}_{j}^{k,(l)} + \textbf{b}_1 )(c)  \nonumber\\
\eta(\{\textbf{H}_{j}^{k,(l)}:x_{i,j}^k=1\}) = \sum_{c=1}^C \alpha_{c,k} \textbf{W}_{2,c} \cdot \sum_{j\in N(i,k)} \textbf{H}_{j}^{k,(l)}
\label{eq:message_passing}
\end{align}
\noindent where $\alpha_{c,k}$ represents the learned weight of the $k$-th type of user behavior from the projected $c$-th latent dimension. $N(i,k)$ denotes the neighboring item nodes connected with user $u_i$ under behavior type of $k$ in the interaction graph $G$. $\textbf{W}_1 \in \mathbb{R}^{C\times d}$ and $\textbf{b}_1 \in \mathbb{R}^{C} $ are learn hyperparameters. We define $\delta(\cdot)$ as ReLU activation function. In our embedding process, we aggregate embeddings from different latent dimensions with weight $\alpha_{c,k}$ and transformation parameter $\textbf{W}_{2,c} \in \mathbb{R}^{d\times d}$. The message passing procedure for the target item node $v_j$ from its adjacent user nodes $N(j,k)$ under the behavior type of $k$ is conducted in an analogous way in Eq~\ref{eq:message_passing}.

\subsection{Message Aggregation Layer}
After performing the propagation of type-specific behavior embeddings between user and item, we propose to aggregate representations from different behavior types, by exploiting the underlying dependencies. In \model\ framework, our message aggregation layer is built upon the attentional neural mechanism. In particular, we first define two transformation weight matrices $\textbf{Q}$ and $\textbf{K}$ for embedding projection between different behavior types $k$ and $k'$. The explicit relevance score between type-specific behavior embeddings is represented as $\beta_{k,k'}$, which is formally calculated as follows:
\begin{align}
    \beta_{k,k'}^s = [(\textbf{Q}^s \textbf{H}^{k,(l)}_{i\leftarrow})^\top \cdot (\textbf{K}^s \textbf{H}^{k',(l)}_{i\leftarrow}) ]/\sqrt{d/S}
\end{align}
\noindent We perform the embedding projection process with multiple latent spaces ($s\in S$), to enable the behavior dependency modeling from different hidden dimensions. $\textbf{Q}^s \in\mathbb{R}^{\frac{d}{S}\times d}$ and $\textbf{K}^s\in\mathbb{R}^{\frac{d}{S}\times d}$ correspond to the transformation matrices of $s$-th projection space. We further apply the softmax function on $\beta_{k,k'}^s$. Then, we recalibrate the type-specific behavior embedding by concatenating representations from different learning subspaces with the following operation:
\begin{align}
\tilde{\textbf{H}}_{i\leftarrow}^{k,(l)} = \xi(\textbf{H}_{i\leftarrow}^{k,(l)}) = \Big ( \mathop{\Bigm|\Bigm|}\limits_{s=1}^S \sum_{k'=1}^K \beta_{k,k'}^s \textbf{V}^s\cdot \textbf{H}_{i\leftarrow}^{k',(l)} \Big ) \oplus \textbf{H}_{i\leftarrow}^{k,(l)} \nonumber
\end{align}
\noindent where $\mathop{\Bigm|\Bigm|}$ represents the concatenation operation and $\textbf{V}_s  \in\mathbb{R}^{\frac{d}{S}\times d}$ denotes the transformation matrix. We define the propagated recalibrated embedding as $\tilde{\textbf{H}}_{i}^{k,(l)}$. To preserve the original type-specific behavioral patterns, the element-wise addition is utilized between the original embedding $\textbf{H}_{i\leftarrow}^{k,(l)}$ and recalibrated representation $\tilde{\textbf{H}}_{i\leftarrow}^{k,(l)}$, \ie, $\hat{\textbf{H}}_{i\leftarrow}^{k,(l)}=\tilde{\textbf{H}}_{i\leftarrow}^{k,(l)} \oplus \textbf{H}_{i\leftarrow}^{k,(l)}$, where $\hat{\textbf{H}}_{i\leftarrow}^{k,(l)}$ is the updated embedding propagated through the connection edge of $k$-th behavior type.

To fuse the behavior type-specific representations during the embedding propagation process, we develop our message aggregation layer with the following functions as:
\begin{align}
    \textbf{H}_{i\leftarrow}^{(l)}=\psi(\{\hat{\textbf{H}}_{i\leftarrow}^{k,(l)}:k=[1,2,...,K]\}) \nonumber\\
    \textbf{H}_{j\leftarrow}^{(l)}=\psi(\{\hat{\textbf{H}}_{j\leftarrow}^{k,(l)}:k=[1,2,...,K]\})
\end{align}
\noindent where $\psi(\cdot)$ represents the message aggregation function. To fuse user/item embeddings from different behavior types, we aim to identify the importance score of individual behavior type-specific representation in assisting the recommendation phase between users and items. To achieve this goal, we feed $\hat{\textbf{H}}_{i\leftarrow}^{k,(l)}$ into a feed-forward neural network to calculate the importance weights as follows (take the user side as example):
\begin{align}
    \gamma_{k}&=\textbf{w}_2^\top\cdot\delta(\textbf{W}_3 \hat{\textbf{H}}_{i\leftarrow}^{k,(l)} + \textbf{b}_2)+b_3\nonumber\\
    \hat{\gamma}_k&=\frac{\exp{\gamma_k}}{\sum_{k'=1}^{K}\exp{\gamma_{k'}}}
\end{align}
\noindent where $\gamma_k$ represents the intermediate values which are fed into a softmax function to generate the importance weight $\hat{\gamma}_k$. In addition, $\delta(\cdot)$ denotes the ReLU activation function. $\textbf{W}_1\in\mathbb{R}^{d'\times d}$ and $\textbf{w}_2 \in \mathbb{R}^{d'}$ represents trainable transformation matrices. $\textbf{b}_1\in\mathbb{R}^{d'}$ and $b_2\in\mathbb{R}$ are bias terms. $d'$ denotes the embedding dimension of hidden layer. After obtaining the weight $\hat{\gamma}_k$ corresponding to behavior type of $k$, the embedding aggregation process is performed as: $\textbf{H}^{(l+1)}_i=\sum_{k=1}^K\hat{\gamma}_k\hat{\textbf{H}}_{i\leftarrow}^{k,(l+1)}$ and $\textbf{H}^{(l+1)}_j=\sum_{k=1}^K\hat{\gamma}_k\hat{\textbf{H}}_{j\leftarrow}^{k,(l+1)}$, where $\textbf{H}^{(l+1)}_i$ and $\textbf{H}^{(l+1)}_j$ serve as the input user/item embedding for $(l+1)$-th graph layer.

Given the generated graph structure of user-item interactions, we learn the high-order multi-behavioral relations over $G=\{U,V,E\}$ by stacking multiple information propagation layers. The embedding propagation between the $(l)$-th and $(l+1)$-th graph layers can be formally represented as below:
\begin{align}
    \textbf{H}^{(l+1)}_{i}&=\psi(\xi(\eta(\{\textbf{H}^{(l)}_j:x_{i,j}^k=1\}))\nonumber\\
    \textbf{H}^{(l+1)}_{U}&=\sum_{k=1}^K\hat{\gamma}_k\cdot\text{MH-Att}(\textbf{X}^k \sum_{c=1}^C \alpha_{c,k} \cdot \textbf{H}_{V}^{k,(l)} \cdot \textbf{W}_{2,c})
\end{align}
\noindent where $\textbf{H}^{(l+1)}_{(i)} \in \mathbb{R}^{I\times d}$ denotes the embeddings of all users for the $(l+1)$-th graph layer. Given the behavior type of $k$, the adjacent relations are represented as $\textbf{X}^k\in \mathbb{R}^{I\times J}$. Similarly, embeddings of items ($v_j \in V$) can be generated based on the above propagation and aggregation operations.

\begin{algorithm}[t]
    \small
	\caption{Training Process of \model}
	\label{alg:train}
	\LinesNumbered
	\KwIn{adjacent tensor $\textbf{X}\in\mathbb{R}^{I\times J\times K}$, $0$-order node features $\bar{\textbf{E}}^{(0)}$, maximum GNN layer $L$, training sampling number $S$, maximum epoch number $N$, regularization weight $\lambda$}
	\KwOut{trained model parameters $\mathbf{\Theta}$}
	Initialize all parameters $\mathbf{\Theta}$\\
	\For{$n=1$ to $N$}{
	    Randomly sample seed nodes $\mathbb{U}$, $\mathbb{V}$\\
    	Get $\textbf{H}^{(0)}$ from $\bar{\textbf{H}}^{(0)}$ for $u_i$ in $\mathbb{U}$ and $v_j$ in $\mathbb{V}$\\
    	\For{$l=1$ to $L$}{
    	    \For{each $u_i$ in $\hat{\mathbb{U}}$, $v_j$ in $\hat{\mathbb{V}}$ and $k=1$ to $K$}{
    	        Construct type-specific behavior message $\textbf{H}^k$\\
    	        Recalibrate the message and get $\hat{\textbf{H}}^k$\\
    	        Acquire the aggregated embedding $\textbf{H}^{(l)}$\\
        	}
    	}
    	$\mathcal{L}=\lambda\|\mathbf{\Theta}_\text{F}^2\|$\\
    	\For{each $u_i$ in $\hat{\mathbb{U}}$}{
    	    Sample $S$ positive and $S$ negative items $v_{p_s}$ and $v_{n_s}$ from $\hat{\mathbb{V}}$\\
    	    \For{each $v_{p_s}$ and ${v_{n_s}}$} {
    	        Calculate $\text{Pr}_{i,j}$ with multi-order matching\\
    	        $\mathcal{L}+=\max(0,1-\text{Pr}_{i,p_s}+\text{Pr}_{i,n_s})$
    	    }
    	}
    	Optimize $\mathbf{\Theta}$ using Adam with loss $\mathcal{L}$\\
    }
    return $\mathbf{\Theta}$
	
\end{algorithm}

\subsection{Model Optimization of \model}
To optimize our \model\ model and infer the hyperparameters, we perform the learning process with the pairwise loss which has been widely used in item recommendation task. Specifically, for each user $u_i$ in the mini-batch training, we define the positive interacted items (\ie, $(v_{p_1}, v_{p_2},...,v_{p_S})$) of user $u_i$ as $S$. For generating negative instances, we randomly sample $S$ non-interacted items $(v_{n_1}, v_{n_2},...v_{n_S})$ of user $u_i$. Given the sampled positive and negative instances, we define our loss function as follows:
\begin{align}
    \mathcal{L}=\sum_{i=1}^I\sum_{s=1}^S\max(0,1-\text{Pr}_{i,p_s}+\text{Pr}_{i,n_s})+\lambda\|\mathbf{\Theta}\|_{\text{F}}^2
\end{align}
\noindent we incorporate the regularization term $|\mathbf{\Theta}\|_{\text{F}}^2$ with the parameter $\lambda$. The learnable parameters are denoted as $\mathbf{\Theta}$. The training process of our model is elaborated in Algorithm~\ref{alg:train}.



\section{Evaluation}
\label{sec:eval}

In this section, we conduct experiments to evaluate our proposed \emph{\model} method by comparing it with state-of-the-art baselines on real-world dataests.



\subsection{Experimental Settings}

\subsubsection{\bf Data Description}
\label{sec:data}
We evaluate our \emph{\model} on three datasets collected from MovieLens, Yelp and Taobao platforms. The statistical information of them is shown in Table~\ref{tab:data}.

\begin{itemize}[leftmargin=*]
\item \textbf{MovieLens Data}\footnote{https://grouplens.org/datasets/movielens/10m/}. It is a benchmark dataset for performance evaluation of recommender systems. In this data, we differentiate users' interaction behaviors over items in terms of the rating scores, \ie, $r_{i,j} \leq 2$: \emph{dislike} behavior. (2) 2$<r_{i,j}<$4. \emph{neutral} behavior. (3) $r_{i,j} > 4$: \emph{like} behavior.

\item \textbf{Yelp Data}\footnote{https://www.yelp.com/dataset/download}. This dataset is collected from the public data repository from Yelp platform. Besides the users' rating data, this data also contains the tip behavior if user gives a tip on his/her visited venues. Ratings are also mapped into three interactions types with the same partition strategy in MovieLens. Similarly, like behavior is the our target and other auxiliary behaviors are \{tip, neutral, dislike\}.

\item \textbf{Taobao Data}\footnote{https://tianchi.aliyun.com/dataset/dataDetail?dataId=649\&userId=1}. This dataset contains different types of user behaviors from Taobao platform, \ie, page view, add-to-cart, add-to-favorite and purchase.

\end{itemize}

\begin{table}[t]
    \caption{Statistics of the experimented datasets}
    \vspace{-0.1in}
    \label{tab:data}
    \centering
    \footnotesize
	\setlength{\tabcolsep}{0.6mm}
    \begin{tabular}{ccccc}
        \toprule
        Dataset&User \#&Item \#&Interaction \#&Interactive Behavior Type\\
        \midrule
        Yelp&19800&22734&$1.4\times 10^6$&\{Tip, Dislike, Neutral, Like\}\\
        ML10M&67788&8704&$9.9\times 10^6$&\{Dislike, Neutral, Like\}\\
        Taobao & 147894 & 99037 & $7.6\times 10^6$ & \{Page View, Favorite, Cart, Purchase\}\\
        \hline
    \end{tabular}
\vspace{-0.2in}
\end{table}

\subsubsection{\bf Evaluation Metrics}
We utilize two metrics \textit{Hit Ratio (HR@$N$)} and \textit{Normalized Discounted Cumulative Gain (NDCG@$N$)} for evaluation. The higher HR and NDCG scores represent more accurate recommendation results. We sample each positive instance with 99 negative instances from users' interacted and non-interacted items, respectively.

\subsubsection{\bf Baselines}
We consider the following baselines:
\begin{itemize}[leftmargin=*]
\item \textbf{BiasMF}~\cite{koren2009matrix}: This method enhances the matrix factorization framework with the consideration of user and item bias. \\\vspace{-0.12in}
\item \textbf{DMF}~\cite{xue2017deep}: It integrates the matrix factorization and neural network to project users and items into embeddings. \\\vspace{-0.12in}
\item \textbf{NCF}~\cite{he2017neural}: It augments the collaborative filtering with deep neural networks. We consider three variants with different feature modeling methods, \ie, i) NCF-N: fusing the matrix factorization and multi-layer perceptron; ii) NCF-G: performing fixed element-wise product on user and item embeddings; iii) NCF-M: using multi-layer perceptron to model the interaction between user's and item's features. \\\vspace{-0.12in}
\item \textbf{AutoRec}~\cite{sedhain2015autorec}: It is based on the autoencoder paradigm for embedding generation in collaborative filtering with the reconstruction objective in the output space. \\\vspace{-0.12in}
\item \textbf{CDAE}~\cite{wu2016collaborative}: In CDAE, a denoising autoencoder model is developed to learn latent representations of users and items with the incorporation of non-linearities. \\\vspace{-0.12in}
\item \textbf{NADE}~\cite{zheng2016neural}: It is a feed-forward neural autoregressive framework with parameter sharing strategy between different ratings of users, to improve collaborative filtering. \\\vspace{-0.12in}
\item \textbf{CF-UIcA}~\cite{du2018collaborative}: It performs autoregression to capture correlations between users and items for collaborative filtering. \\\vspace{-0.12in}
\item \textbf{NGCF}~\cite{wang2019neural}: It is a graph neural collaborative filtering model to project users and items into latent representations over the structure of user-item interaction graph. The embedding propagation is performed across graph layers. \\\vspace{-0.12in}
\item \textbf{NMTR}~\cite{gao2019neural}: It is a multi-task learning framework to correlate the prediction of different types of user behaviors. In NMTR framework, a shared embedding layer is designed for different behavior types of interactions. \\\vspace{-0.12in}
\item \textbf{DIPN}~\cite{guo2019buying}: This method is on the basis of attention mechanism and recurrent neural network to aggregate signals from users' browse and purchase behaviors.
\end{itemize}

\subsubsection{\bf Parameter Settings}
We implement our \emph{\model} model with TensorFlow and the model is optimized using the Adam optimizer during the training phase. We set the dimension of embeddings as 16 and the number of latent dimensions in our memory neural module as 8. In addition, the batch size and learning rate in our model is set as 32 and $1e^{-3}$. The decay rate of 0.96 is applied during the learning process.

\begin{table}[t]
	\caption{Performance Comparison in terms of \textit{HR@$10$} and \textit{NDCG@$10$}.}
	\vspace{-0.1in}
	\centering
    \footnotesize
	\setlength{\tabcolsep}{0.6mm}
	\begin{tabular}{|c|c|c|c|c|c|c|c|c|}
		\hline
		\multirow{2}{*}{Model}&
		\multicolumn{2}{c|}{MovieLens Data}&\multicolumn{2}{c|}{Yelp Data}&\multicolumn{2}{c|}{Taobao Data}\\
		\cline{2-7}
		&~~HR~~ & NDCG & ~~HR~~ & NDCG & ~~HR~~ & NDCG \\
		\hline
		\hline
		BiasMF  & 0.767 & 0.490 & 0.755 & 0.481 & 0.262 & 0.153 \\
		\hline
		DMF     & 0.779 & 0.485 & 0.756 & 0.485 & 0.305 & 0.189 \\
		\hline
		NCF-M   & 0.757 & 0.471 & 0.714 & 0.429 & 0.319 & 0.191 \\
		\hline
		NCF-G   & 0.787 & 0.502 & 0.755 & 0.487 & 0.290 & 0.167 \\
		\hline
		NCF-N   & 0.801 & 0.518 & 0.771 & 0.500 & 0.325 & 0.201 \\
		\hline
		AutoRec & 0.658 & 0.392 & 0.765 & 0.472 & 0.313 & 0.190 \\
		\hline
		CDAE    & 0.659 & 0.392 & 0.750 & 0.462 & 0.329 & 0.196 \\
		\hline
		NADE    & 0.761 & 0.486 & 0.792 & 0.499 & 0.317 & 0.191 \\
		\hline
		CF-UIcA & 0.778 & 0.491 & 0.750 & 0.469 & 0.332 & 0.198 \\
		\hline
		NGCF    & 0.790 & 0.508 & 0.789 & 0.500 & 0.302 & 0.185 \\
		\hline
		NMTR    & 0.808 & 0.531 & 0.790 & 0.478 & 0.332 & 0.179 \\
		\hline
		DIPN    & 0.791 & 0.500 & 0.811 & 0.540 & 0.317 & 0.178 \\
		\hline
		\emph{\model} & \textbf{0.857} & \textbf{0.575} & \textbf{0.848} & \textbf{0.559} & \textbf{0.424} & \textbf{0.249}\\
		\hline
	\end{tabular}
	\label{tab:target_behavior_results}
	\vspace{-0.1in}
\end{table}

\subsection{Performance Comparison}
In this section, we evaluate all approaches in predicting like behaviors of user over different items on different real-world datasets. We report HR@10 and NDCG@10 in Table~\ref{tab:target_behavior_results}. We can observe that \emph{\model} outperforms all baselines on various datasets in terms of both HR@10 and NDCG@10. For example, \emph{\model} achieves relatively 6.06\% and 8.34\% improvements in terms of HR, and 8.28\% and 15.0\% improvements in terms of NDCG, over NMTR and DIPN, respectively on MovieLens dataset. Such significant performance gain could be attributed to the explicitly modeling of inter-dependencies among different types of user-item interactions. The best performance is followed by NMTR which defines the correlations between different types of user behaviors in a cascaded way. In contrast, our \emph{\model} model automatically captures the implicit behavior dependencies based on our developed graph neural networks. Furthermore, the performance gap between \emph{\model} and the state-of-the-art graph neural collaborative filtering model--NGCF, justifies the advantage of incorporating the cross-interaction collaborative relations into the graph neural architecture for user's preference encoding.

In addition, we can observe that recommendation techniques (\ie, NMTR and DIPN) with the consideration of multi-behavioral patterns improve the performance as compare to other types of baselines which fail to differentiate interaction relationships between user and item. This points to the positive effect of incorporating multi-typed interactive patterns in the embedding function of recommender systems. Among other baselines, the graph neural network-based model (\ie, NGCF) achieves better performance in most cases, which justifies the rationality of encoding the collaborative signals by conducting the embedding propagation between neighboring nodes based on user-item graph-structural relations.

We further measure the ranking quality of top-$N$ recommended items with varying $N$ from 1 to 9 with an increment of 2. The evaluation results (measured by HR@$N$ and NDCG@$N$) on Yelp data are presented in Table~\ref{tab:vary_k}. We can observe that \emph{\model} achieves the best performance under different values of $N$, which suggests that \emph{\model} assigns higher score to the user's interested item in the top-ranked list and hit the ground truth at top positions.

\begin{table}[h]
	\caption{Ranking performance evaluation on Yelp dataset with varying Top-\textit{N} value in terms of \textit{HR@N} and \textit{NDCG@N}}
	\vspace{-0.10in}
	\centering
    \footnotesize
	\setlength{\tabcolsep}{0.6mm}
	\begin{tabular}{|c|c|c|c|c|c|c|c|c|c|c|}
		\hline
		\multirow{2}{*}{Model} &\multicolumn{5}{c|}{HR}&\multicolumn{5}{c|}{NDCG}\\
		\cline{2-11}
		&@1&@3&@5&@7&@9&@1&@3&@5&@7&@9\\
		\hline
		\hline
		BiasMF
		& 0.287 & 0.474 & 0.626 & 0.714 & 0.741 & 0.287 & 0.378 & 0.432 & 0.461 & 0.474 \\
		\cline{2-11}
		\hline
        NCF-N
        &0.260&0.481&0.604&0.695&0.742&0.260&0.396&0.444&0.477&0.492\\
        \hline
        AutoRec
        &0.228&0.455&0.586&0.684&0.732&0.228&0.362&0.410&0.449&0.462\\
        \hline
        NADE
        &0.265&0.508&0.642&0.720&0.784&0.265&0.402&0.454&0.478&0.497 \\
        \hline
        CF-UIcA
        & 0.235&0.449&0.576&0.659&0.731 &0.235&0.360&0.412&0.440&0.463 \\
        \hline
        NMTR
        & 0.214&0.466&0.610&0.700&0.762 &0.214&0.360&0.419&0.450&0.469 \\
        \hline
        \emph{\model}
        & \textbf{0.320} & \textbf{0.590} & \textbf{0.700} & \textbf{0.784} & \textbf{0.831} & \textbf{0.320} & \textbf{0.473} & \textbf{0.519} & \textbf{0.542} & \textbf{0.558} \\
		\hline
	\end{tabular}
	\label{tab:vary_k}
	\vspace{-0.1in}
\end{table}

\subsection{Component Ablation Evaluation}

\begin{figure}
	\centering
	\vspace{-0.1 in}
	\subfigure[][ML-HR]{
		\centering
		\includegraphics[width=0.102\textwidth]{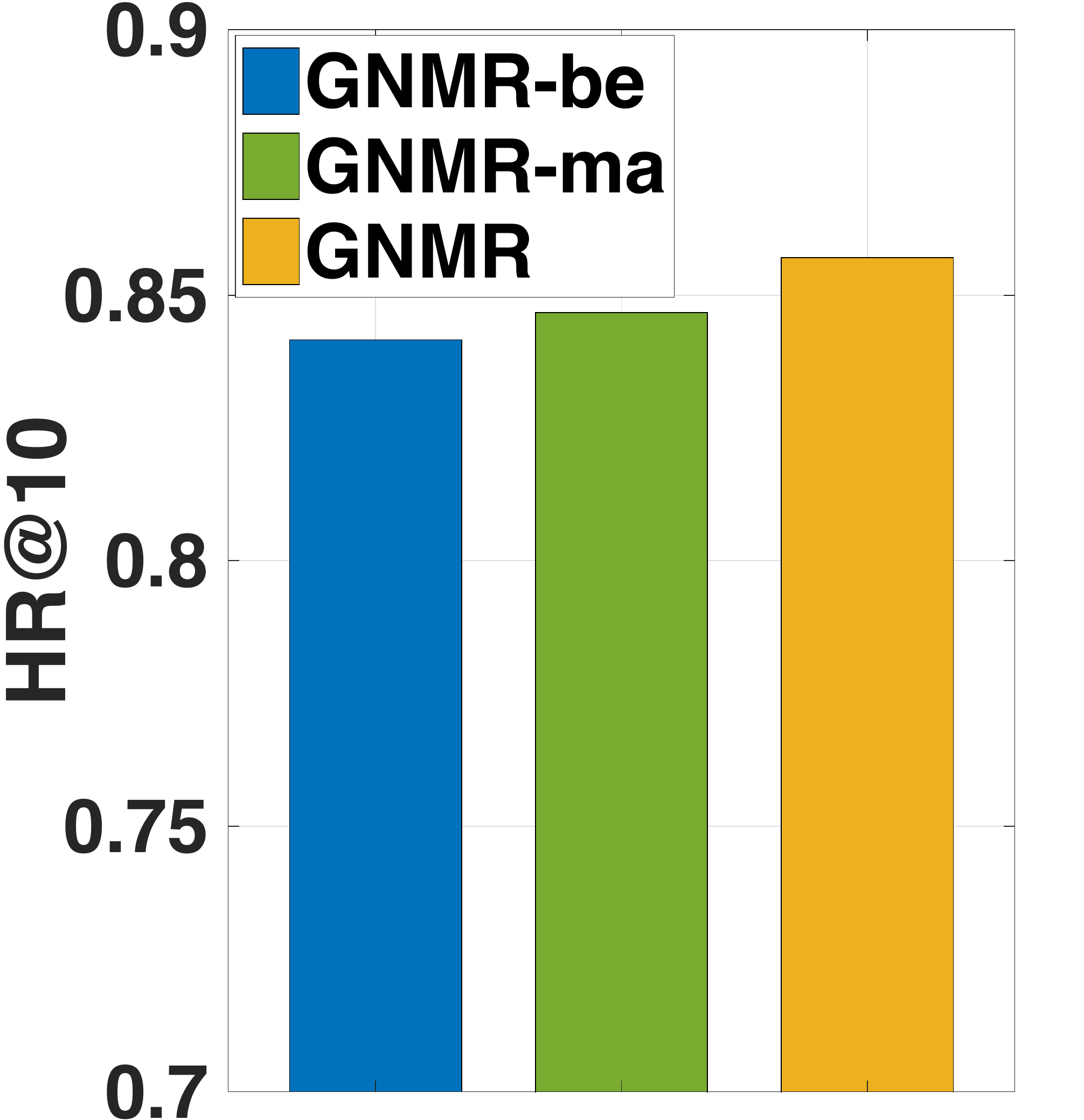}
		\label{fig:ablation_yelp_HR}
	}
	\subfigure[][ML-NDCG]{
		\centering
		\includegraphics[width=0.1\textwidth]{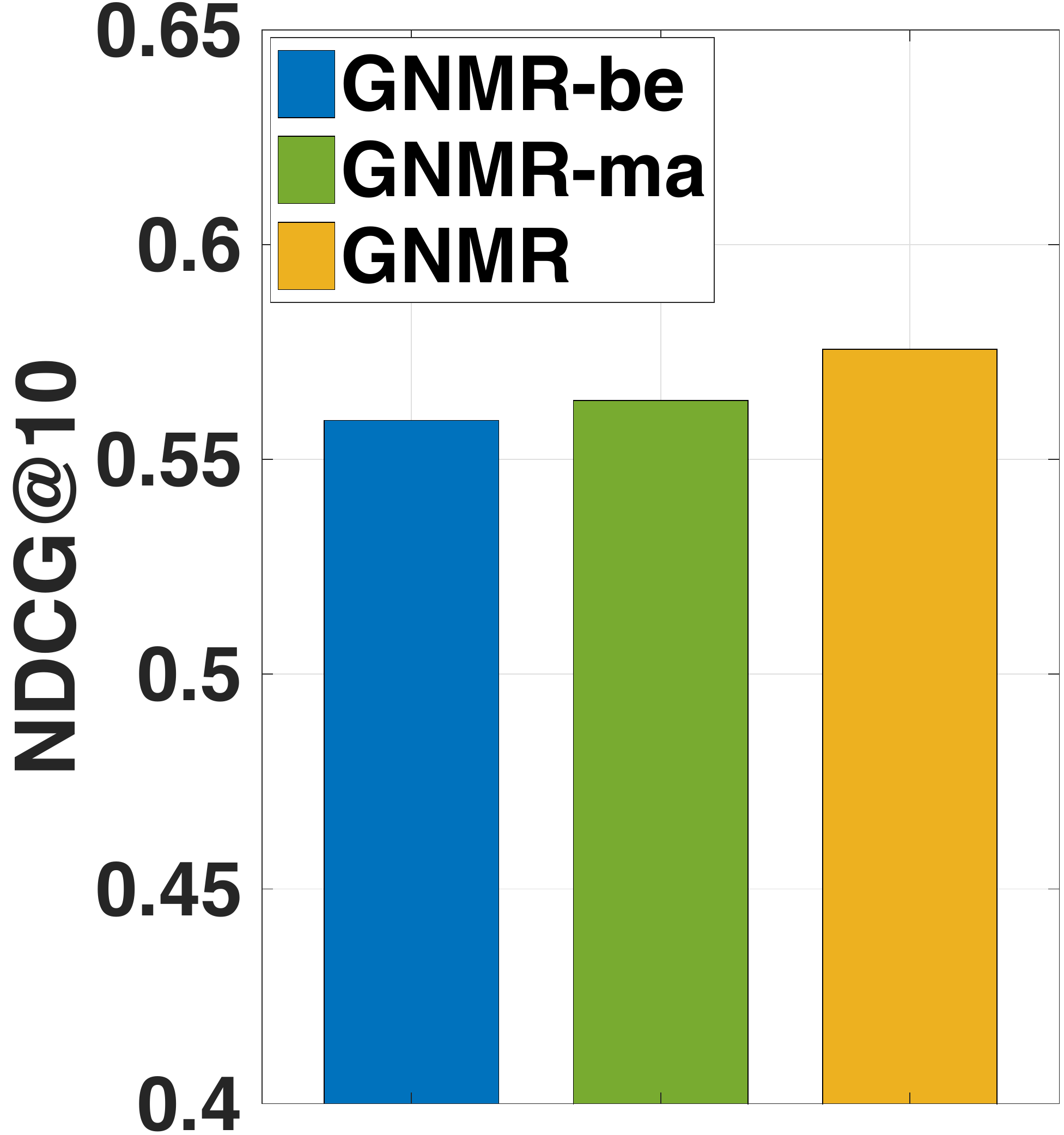}
		\label{fig:ablation_yelp_NDCG}
	}
	\subfigure[][Yelp-HR]{
		\centering
		\includegraphics[width=0.1\textwidth]{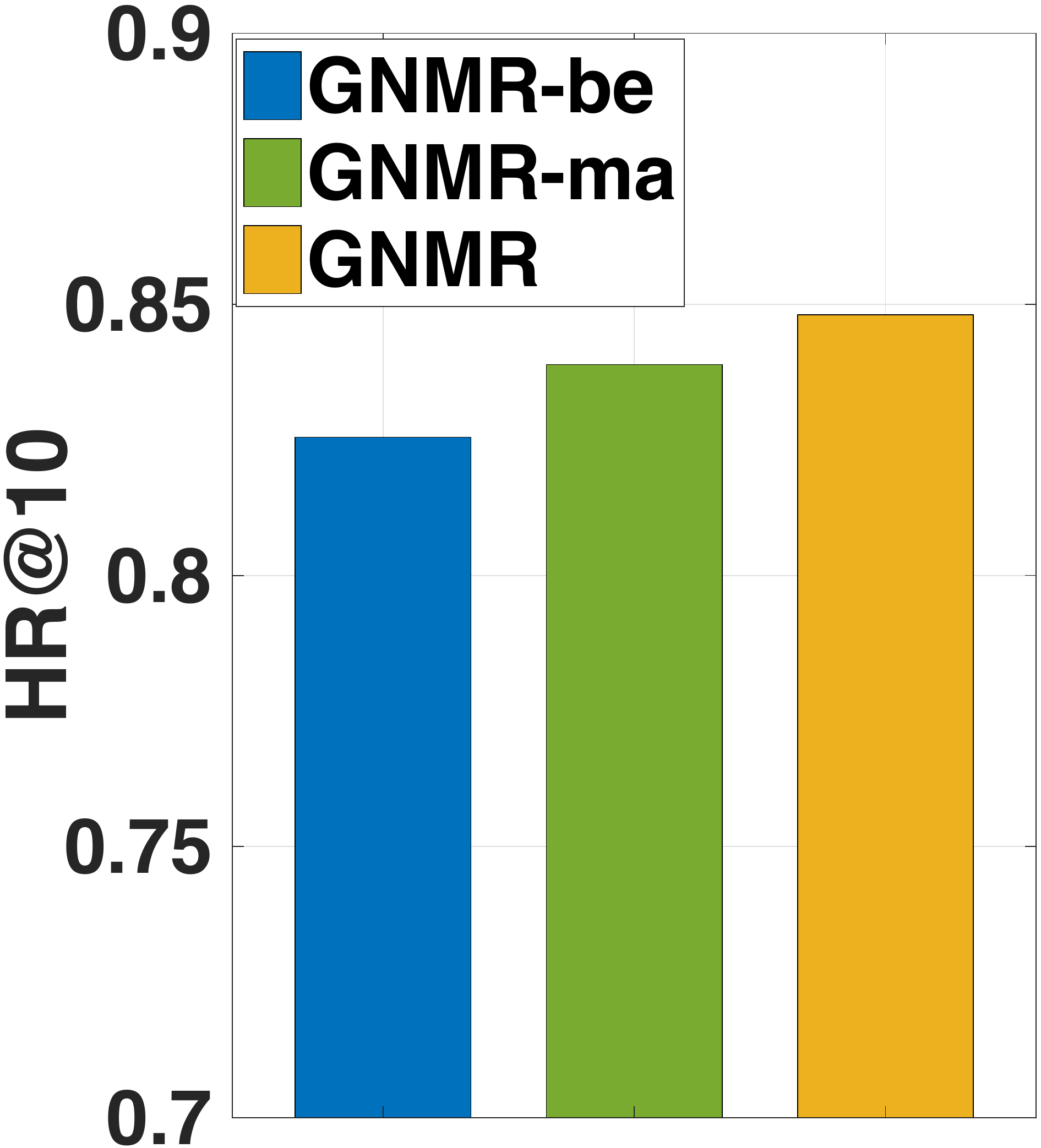}
		\label{fig:ablation_ml10m_HR}
	}
	\subfigure[][Yelp-NDCG]{
		\centering
		\includegraphics[width=0.1\textwidth]{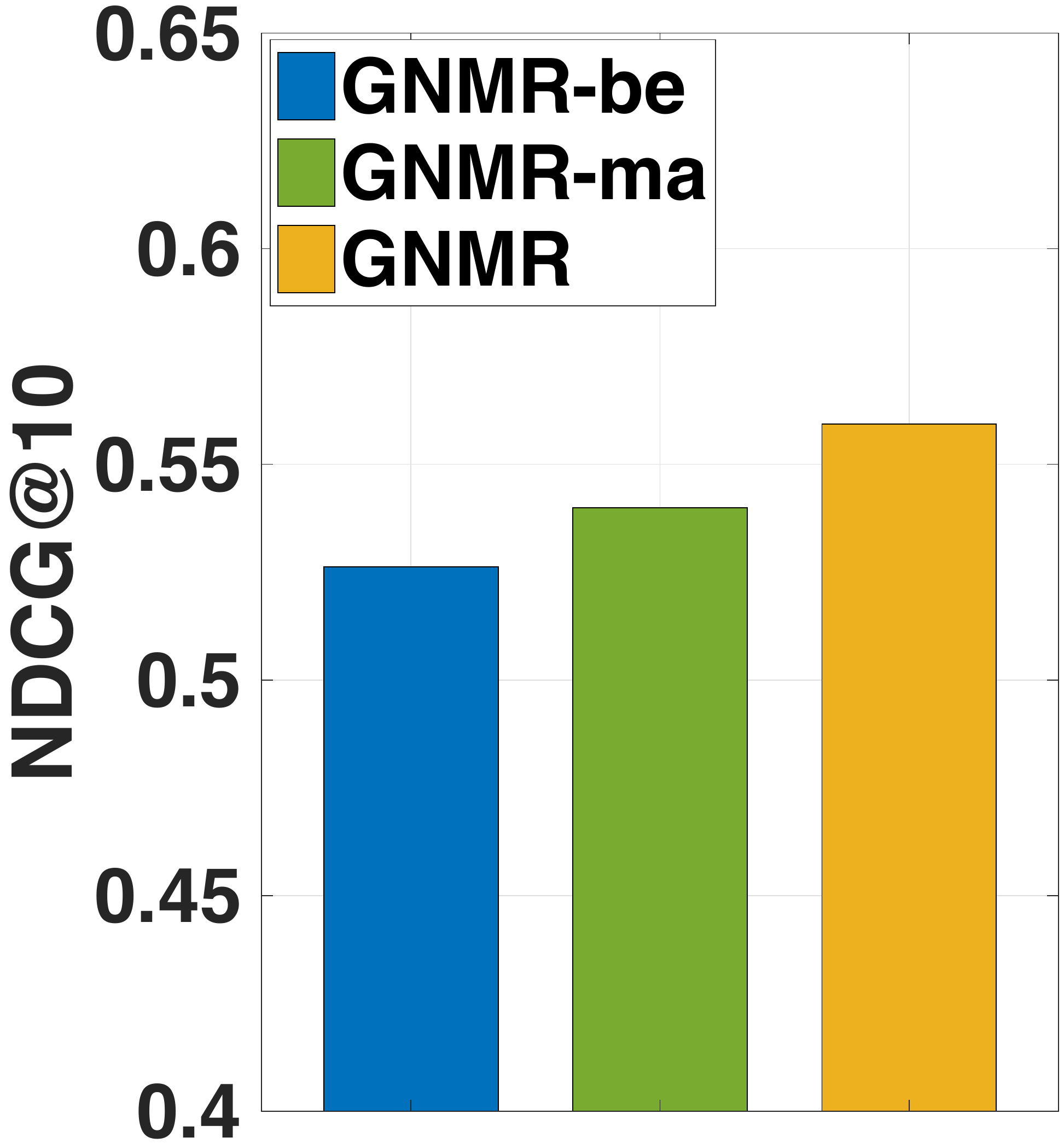}
		\label{fig:ablation_ml10m_NDCG}
	}
	\vspace{-0.05in}
	\caption{Ablation study of \emph{\model} framework on Yelp and MovieLens data in terms of \textit{HR@10} and \textit{NDCG@10}.}
	\label{fig:ablation}
	\vspace{-0.2in}
\end{figure}

We further conduct experiments to evaluate the component-wise effect of our proposed \emph{\model} framework. Evaluation results of several designed model variants are shown in Figure~\ref{fig:ablation}.\\\vspace{-0.1in}

\noindent \textbf{Type-specific Behavior Embedding Component}.
We first design a contrast method \emph{\model}-be without the type-specific behavior embedding layer, to generate the propagated feature embedding of individual type of user behavior during the message passing process. The performance gap between \emph{\model}-be and \emph{\model} indicates the effectiveness of our attention-based projection layer to capture the type-specific behavioral features.\\\vspace{-0.15in}

\noindent \textbf{Message Aggregation Component}. Another dimension of model ablation study aims to evaluate the efficacy of our message aggregation layer. \emph{\model}-ma is the variant of our framework, in which the behavior dependency modeling component is removed in our graph-structured embedding propagation. From Figure~\ref{fig:ablation}, we observe that \emph{\model} outperforms \emph{\model}-ma, which shows that capturing the type-wise behavior dependencies is helpful for learning relation-aware collaborative signals and improving the recommendation performance.

\begin{table}[h]
    \vspace{-0.05in}
	\caption{Performance evaluation of our \emph{\model} with the aggregation of different types of behavioral patterns.}
	\vspace{-0.10in}
	\centering
    \footnotesize
	\setlength{\tabcolsep}{0.6mm}
	\begin{tabular}{|c|c|c|c|c|c|c|}
		\hline
		Dataset & Metric & w/o dislike & w/o neutral & w/o like & only like & \emph{\model} \\
		\cline{1-7}
		\multirow{2}{*}{MovieLens} & HR & 0.834 & 0.816 & 0.838 & 0.835 & \textbf{0.857} \\
		\cline{2-7}
		& NDCG & 0.549	& 0.532 & 0.559 & 0.559 & \textbf{0.575} \\
		\cline{1-7}
		Dataset & Metric & w/o tip & w/o dislike & w/o neutral & only like & \emph{\model} \\
		\cline{1-7}
		\multirow{2}{*}{Yelp} & HR & 0.837 & 0.833 & 0.831 & 0.821 & \textbf{0.848} \\
		\cline{2-7}
		& NDCG & 0.535 & 0.542 & 0.532 & 0.527 & \textbf{0.559} \\
		\hline
	\end{tabular}
	\label{tab:behavior_type}
	\vspace{-0.15in}
\end{table}

\subsection{Performance with Various Behavior Types}
We also perform experiments to evaluate the recommendation accuracy with the consideration of different user-item relations from multi-typed user behaviors on both MovieLens and Yelp. Particularly, for MovieLens, we first design three variants: w/o dislike, w/o neutral, w/o like indicate that we do not include the dislike, neutral and like behavior respectively into our multi-behavior enhanced graph-based recommendation framework. We further introduce another model variant (only like) which merely relies on the target type of behaviors for making recommendation. We apply similar settings to Yelp data with four variants of our \emph{\model} method. Evaluation results in terms of HR@10 and NDCG@10 are listed in Table~\ref{tab:behavior_type}. As we can see, the improvement is quite obvious when we consider more diverse behavior types for multi-behavioral knowledge integration in our graph neural network. In addition, compared with the model variant which only considers the target type of behavior data (\ie, like behavior) for prediction, we find that our multi-behavior enhanced recommender system \emph{\model} do help to improve the performance, which indicates the positive effect of incorporating auxiliary knowledge from other types of user-item interaction behaviors into the embedding function of recommendation model.


\subsection{Impact of Model Depth in \emph{\model} Framework}
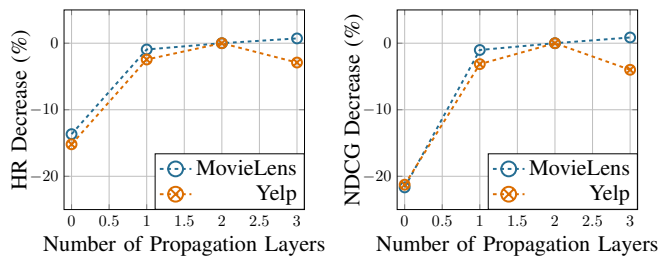
\begin{figure}
\vspace{-0.1in}
    \centering
    \begin{adjustbox}{max width=1.0\linewidth}
    \begin{tikzpicture}
\begin{axis}[
    xlabel={Number of Propagation Layers},
    ylabel={HR Decrease (\%)},
    xmin=-0.1,xmax=3.1,
    ymin=-25,ymax=5,
    legend columns=1,
    legend cell align=right,
    grid=both,
    every axis plot/.append style={ultra thick},
    every tick label/.append style={scale=1.3},
    label style={scale=1.8},
    legend style={
        nodes={scale=1.5, transform shape},
        legend image post style={scale=1.5},
        },
    legend style={at={(1,0)},anchor=south east},
    every axis plot post/.append style={
        every mark/.append style={scale=2}
    }
]
\addplot[color={rgb:blue,4;green,2;yellow,1}, mark=o, dashed, mark options={solid}]
table[x=para, y=yelp_hr] {gnnLayer-buy.txt};
\addplot[color={rgb:red,4;green,1;yellow,2}, mark=otimes, dashed, mark options={solid}]
table[x=para, y=ml10m_hr] {gnnLayer-buy.txt};
\legend{\large MovieLens, \large Yelp};
\end{axis}
\end{tikzpicture}

\begin{tikzpicture}
\begin{axis}[
    xlabel={Number of Propagation Layers},
    ylabel={NDCG Decrease (\%)},
    xmin=-0.1,xmax=3.1,
    ymin=-25,ymax=5,
    legend columns=1,
    legend cell align=right,
    grid=both,
    every axis plot/.append style={ultra thick},
    every tick label/.append style={scale=1.3},
    label style={scale=1.8},
    legend style={
        nodes={scale=1.5, transform shape},
        legend image post style={scale=1.5},
        },
    legend style={at={(1,0)},anchor=south east},
    every axis plot post/.append style={
        every mark/.append style={scale=2}
    }
]
\addplot[color={rgb:blue,4;green,2;yellow,1}, mark=o, dashed, mark options={solid}]
table[x=para, y=yelp_ndcg] {gnnLayer-buy.txt};
\addplot[color={rgb:red,4;green,1;yellow,2}, mark=otimes, dashed, mark options={solid}]
table[x=para, y=ml10m_ndcg] {gnnLayer-buy.txt};
\legend{\large MovieLens, \large Yelp};

\end{axis}
\end{tikzpicture}
    \end{adjustbox}
    \vspace{-0.3in}
    \caption{Impact study of model depth in terms of \textit{HR@10} and \textit{NDCG@10}.}
    \vspace{-0.2in}
    \label{fig:hyperparam_buy}
\end{figure}

In this subsection, we examine the influence of the depth of \emph{\model} to investigate the effectiveness of stacking multiple information propagation layers. Specifically, we vary the depth of our graph neural networks from 0 (without the message passing mechanism) to 3. \emph{\model}-1 is defined to represent the method with one embedding propagation layer and similar notations for other variants. In Figure~\ref{fig:hyperparam_buy}, the y-axis represents the performance variation percentage compared with \emph{\model}-2, so as to present results of different datasets in the same figure. From evaluation results, we can observe that \emph{\model}-2 and \emph{\model}-3 outperform \emph{\model}-1, which justifies the effective modeling of high-order collaborative effects in our multi-behavior enhanced recommendation scenario. Learning second- or third-order collaborative relations is sufficient to encode the user-item interactive patterns. Further increasing the number of propagation layers is prone to involve noise in our graph-based neural collaborative filtering.


\section{Conclusion}
\label{sec:conclusion}

This paper contributes a new framework, named \model\ for multi-behavior enhanced recommendation through modeling inter-type behavior dependencies under a graph-structured message passing architecture. \model\ effectively aggregates the heterogeneous behavioral patterns of users by performing embedding propagation over user-item interaction graph. Our \model\ could capture the heterogeneous relations among different types of user-item interactions for encoding graph-structured collaborative signals for recommendation. Through experiments on real-world datasets, our proposed \model\ is shown to achieve better performance compared with state-of-the-art baselines. Furthermore, the ablation studies show that \model\ is able to comprehensively leverage the multi-typed behavioral patterns to improve recommendation accuracy. In future, we are interested in exploring the attribute features from user and item side, such as user profile and item textual information, so as to further alleviate the data sparsity problem. More generally, we could extend our \model\ framework to model other heterogeneous relationships (\eg, different social connections or item dependencies) in recommendation.

\section*{Acknowledgments}
We thank the anonymous reviewers for their constructive feedback and comments. This work is supported by National Nature Science Foundation of China (62072188, 61672241), Natural Science Foundation of Guangdong Province (2016A030308013), Science and Technology Program of Guangdong Province (2019A050510010).

\bibliographystyle{abbrv}
\bibliography{refs} 

\end{document}